\documentclass[twocolumn,tighten,times]{main}

\newcommand\aastex{AAS\TeX}
\newcommand\latex{La\TeX}
\newcommand{\lya}{Ly$\alpha$}
\newcommand{\sigmasfr}{$\rm \Sigma_{SFR}$}
\newcommand{\sigmassfr}{$\rm \Sigma_{sSFR}$}
\newcommand{\mstar}{$M_*$}
\newcommand{\ebv}{$E(B-V)$}
\newcommand{\ebvcont}{$E(B-V)_{\rm cont}$}
\newcommand{\fdtb}{$f_{\rm dtb}$}

\graphicspath{{./}{figures/}}
\begin{document}

\title{\lya{} Halo Properties and Dust in the Circumgalactic Medium of z$\sim$2 Star-forming Galaxies}

\author{Zhiyuan Song}
\affiliation{Department of Physics and Astronomy, University of California, Riverside, 900 University Avenue, Riverside, CA 92521, USA}
\email{zhiyuan.song@email.ucr.edu}

\author{Naveen A. Reddy}
\affiliation{Department of Physics and Astronomy, University of California, Riverside, 900 University Avenue, Riverside, CA 92521, USA}

\author{Yuguang Chen}
\affiliation{Cahill Center for Astronomy and Astrophysics, California Institute of Technology, MC249-17, Pasadena, CA 91125, USA}
\affiliation{Department of Physics and Astronomy, University of California Davis, 1 Shields Avenue, Davis, CA 95616, USA}

\author{Alice E. Shapley}
\affiliation{Department of Physics and Astronomy, University of California, Los Angeles, CA 90095, USA}

\author{Saeed Rezaee}
\affiliation{Department of Physics and Astronomy, University of California, Riverside, 900 University Avenue, Riverside, CA 92521, USA}

\author{Andrew Weldon}
\affiliation{Department of Physics and Astronomy, University of California, Riverside, 900 University Avenue, Riverside, CA 92521, USA}

\author{Tara Fetherolf}
\affiliation{Department of Earth and Planetary Sciences, University of California, Riverside, CA 92521, USA}

\author{Alison L. Coil}
\affiliation{CCenter for Astrophysics and Space Sciences, University of California, 9500 Gilman Dr., La Jolla, CA 92093, USA}

\author{Bahram Mobasher}
\affiliation{Department of Physics and Astronomy, University of California, Riverside, 900 University Avenue, Riverside, CA 92521, USA}

\author{Charles C. Steidel}
\affiliation{Cahill Center for Astronomy and Astrophysics, California Institute of Technology, MC249-17, Pasadena, CA 91125, USA}



\begin{abstract}

We present Keck Cosmic Web Imager IFU observations around extended \lya{} halos of 27 typical star-forming galaxies with redshifts $2.0 < z < 3.2$ drawn from the MOSFIRE Deep Evolution Field survey. We examine the average \lya{} surface-brightness profiles in bins of star-formation rate (SFR), stellar mass (\mstar), age, stellar continuum reddening, SFR surface density (\sigmasfr), and \sigmasfr{} normalized by stellar mass (\sigmassfr). The scale lengths of the halos correlate with stellar mass, age, and stellar continuum reddening; and anti-correlate with star-formation rate, \sigmasfr, and \sigmassfr. These results are consistent with a scenario in which the down-the-barrel fraction of \lya{} emission is modulated by the low-column-density channels in the ISM, and that the neutral gas covering fraction is related to the physical properties of the galaxies. Specifically, we find that this covering fraction increases with stellar mass, age, and \ebv; and decreases with SFR, \sigmasfr{} and \sigmassfr. We also find that the resonantly scattered \lya{} emission suffers greater attenuation than the (non-resonant) stellar continuum emission, and that the difference in attenuation increases with stellar mass, age, and stellar continuum reddening, and decreases with \sigmassfr. These results imply that more reddened galaxies have more dust in their CGM.

\end{abstract}



\section{Introduction} \label{sec:intro}
Recent observations indicate that extended \lya{} halos are ubiquitous around high-redshift galaxies, based on both stacked images \citep{Steidel11, Matsuda12, Feldmeier13, Momose14, Momose16, Xue17} as well as individual detections \citep{Wisptzki16, Leclercq17, Erb18, Erb22}. There are several scenarios that might explain diffuse halos of \lya{} emission. These include resonant scattering of \lya{} photons produced in star-forming regions and/or AGN \citep{Meier81, Dijkstra06, Zheng10, Steidel11, Dijkstra12, Orsi12}, \lya{} emission powered by the loss of gravitational energy by inflowing gas \citep{Dijkstra09, Faucher-Giguere10, Goerdt10, Rosdahl12, Lake15}, and \lya{} fluorescence due to a nearby ionizing source unrelated to the galaxy \citep{Adelberger06, Mas-Ribas16}. The prevalence of \lya{} halos irrespective of large-scale environment and the large inferred covering fraction of outflowing gas suggests that, for the most part, these halos reflect the resonant scattering of \lya{} photons originating from the sites of star formation within galaxies \citep{Momose14, Byrohl21, Kikuta23}.

The relation between the sizes of extended \lya{} halos and their host galaxies has been investigated in several studies. Previous studies found that the scale length of the \lya{} halo is positively correlated with the total \lya{} luminosity and UV magnitude, while independent of \lya{} equivalent width \citep{Steidel11, Leclercq17, Xue17}. However, \citet{Momose16} found that the scale length is anticorrelated with \lya{} luminosity and rest-frame equivalent width, while the influence of UV magnitude remains unclear. The link between the \lya{} halo size and the UV magnitude of the host galaxy may indicate that SFR plays an important role in powering extended \lya{} emission, which favors the resonant scattering origin of the \lya{} emission. On the other hand, supernovae and/or stellar winds could also regulate the \lya{} halo by carving low-column-density channels in the interstellar medium (ISM) and the circumgalactic medium (CGM) \citep{Gnedin08, Zackrisson13, Ma16, Kimm19, Ma20, Kakiichi21}. The channels would ease the escape of \lya{} photons at small impact parameters \citep{Reddy16}, hence reducing the scale length. Such competing mechanisms together may explain why \citet{Momose16} found no correlation between UV magnitude and \lya{} halo scale length.

While these studies focused on the UV properties of host galaxies, there has been little investigation into how the sizes and shapes of halos scale with the properties of host galaxies, including stellar mass and reddening. Yet, it is perhaps reasonable to think that these halos, which effectively trace the gas content around galaxies, may depend on the maturity (i.e., stellar mass, age, reddening) of their host galaxies. Many works have shown that the HI covering fraction is a key parameter for \lya{} escape \citep{Kornei10, Hayes11, Wofford13, Borthakur14, RiveraThorsen15, Trainor15, Reddy16, Steidel18, Jaskot19, Reddy22}. Since stellar feedback could modulate the HI covering fraction by creating low-column-density channels in the ISM and the CGM \citep{Gnedin08, Ma16, Kimm19, Ma20, Kakiichi21}, it is important to investigate the impact of the SFR surface density on the escape of \lya{} photons. \citet{Reddy22} also found that the galaxy potential plays an important role on the escape of \lya{} photons. However, measuring these quantities requires deep multiwavelength photometric and spectroscopic observations which were lacking in previous studies \citep[e.g.,][]{Wisptzki16}.

Detecting diffuse \lya{} halos around high redshift galaxies is challenging since it requires high sensitivity and adequate spatial resolution. The state-of-the-art IFU instrument Keck Cosmic Web Imager \citep[KCWI;][]{Morrissey18} was designed to detect such halos around typical star-forming galaxies at $z > 2.0$ \citep{Martin10, Morrissey12, Chen21}. KCWI has a wavelength coverage of 3500 - 5600 \AA{} and a spectral resolution of $R \sim 1800$ (for medium slicer and BL grating configuration) for the blue channel. While the Multi-Unit Spectroscopic Explorer \citep[MUSE;][]{Bacon10} offers a larger field of view, KCWI's unparalleled blue sensitivity enables observations of \lya{} halos at lower redshifts where surface brightness dimming is mitigated and the sky background is low.

In this paper, we investigate the relations between the properties of the \lya{} halos and the physical properties of their host galaxies using KCWI IFU data of a sample of 27 galaxies at redshifts $2.0<z<3.2$. These galaxies are selected from the MOSFIRE Deep Evolution Field \citep[MOSDEF;][]{Kriek15} survey. We also study the relations between dust in the CGM and the physical properties of the galaxies. This paper is structured as follows. The observations and data reduction are described in Section \ref{sec:data}. We discuss the \lya{} surface brightness profiles of the subsamples based on the spectral energy distributions (SED) fitting parameters in Section \ref{sec:composite}, and the reddening of \lya{} photons in the CGM in Section \ref{sec:individual}. In Section \ref{sec:discussion}, the relations between \lya{} halo sizes and physical quantities are discussed. In addition, we also discuss the implications of our results for the dust content in the CGM. Our results are summarized in Section \ref{sec:conclusion}. We use physical distances and assume a $\rm \Lambda CDM$ universe with $\rm \Omega_m = 0.3$, $\rm \Omega_{\Lambda} = 0.7$, and $H_0 = 70 {\rm \ km\ s^{-1}Mpc^{-1}}$.

\section{Observations and Data Reduction} \label{sec:data}

\subsection{MOSDEF Survey} \label{subsec:mosdef}
Our sample was drawn from the MOSDEF survey \citep{Kriek15}, which obtained rest-frame optical spectroscopy of $\sim 1500$ $H$-band selected star-forming galaxies and AGNs in the CANDELS fields \citep[AEGIS, COSMOS, GOODS-N, GOODS-S and UDS;][]{Grogin11, Koekemoer11}. The MOSDEF survey used the MOSFIRE spectrograph \citep{Mclean12} on the Keck I telescope to obtain moderate resolution ($R \sim $ 3000 -- 3600) rest-frame optical spectra ($\sim$ 3700 -- 7000 \AA) at redshifts 1.4 $\lesssim z \lesssim$ 3.8. MOSDEF galaxies were selected based on pre-existing photometric, grism, or spectroscopic redshifts where the strong rest-frame optical lines fall in the $YJHK$ atmospheric transmission windows ($1.37 \leq z \leq 1.70, 2.09 \leq z \leq 2.61, 2.95 \leq z \leq 3.80$). Details on the MOSDEF data reduction are provided in \citet{Kriek15}.

Emission lines were measured from the MOSDEF spectra using a Gaussian function and a linear continuum. The [O\,{\footnotesize{II}}] doublet was fitted with a double Gaussian function, and the H$\alpha$ and [N\,{\footnotesize{II}}] doublet was fit with three Gaussians. Line fluxes and errors were derived by perturbing the spectrum of each object by its error spectrum to generate 1,000 realizations, measuring the line fluxes from each realization, and calculating the average lines fluxes and dispersion. Systematic redshifts were derived using the strongest emission line. We refer the readers to \citet{Kriek15} and \citet{Reddy15} for further details on line flux measurements and slit loss corrections.

\subsection{Sample Selection} \label{subsec:sample}

\begin{figure}[t!]
\includegraphics[width=\columnwidth]{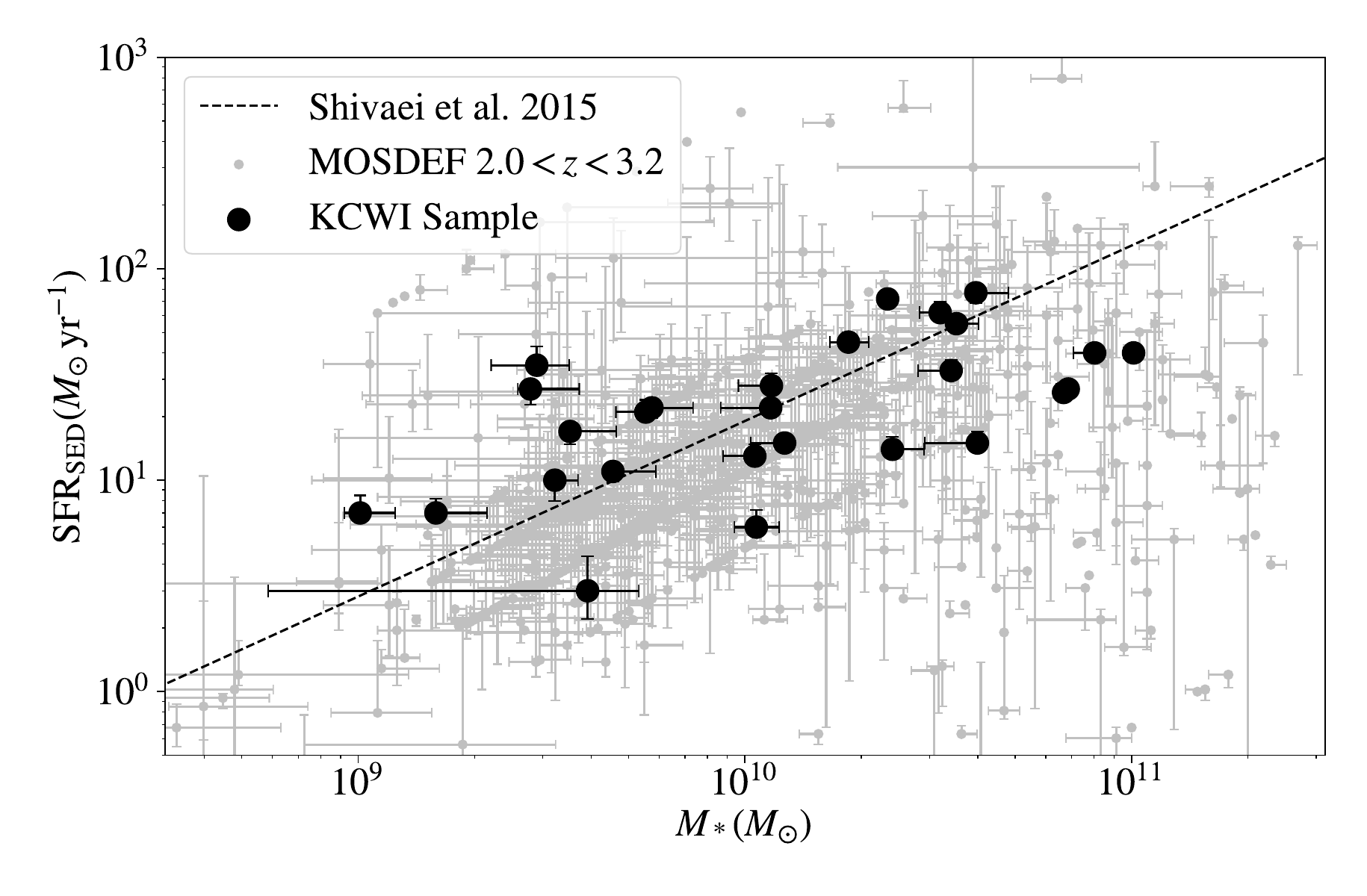}
\caption{SFR vs. $M_*$ for our KCWI sample (black) and MOSDEF galaxies (grey) with $2.0 < z < 3.2$. Both SFR and \mstar{} are derived through SED fitting, which is described in Section \ref{subsec:sed}. The dashed line shows the best-fit linear relation between log(SFR/($M_{\sun}\,{\rm yr^{-1}}$)) and log($M_*/M_{\sun}$) found by \citet{Shivaei15} for MOSDEF star-forming galaxies at $z = 2.09 - 2.61$.\label{fig:sfr_m}}
\end{figure}

Our sample contains 27 star-forming galaxies with redshifts $2.0 < z < 3.2$ in the AEGIS, COSMOS, GOODS-N and GOODS-S fields. We primarily focus on galaxies with detections of H$\alpha$, H$\beta$, [O\,{\footnotesize{III}}], and either a detection or upper limit on [N\,{\footnotesize{II}}]. These galaxies cover the full ranges of stellar mass and SFR for galaxies in MOSDEF survey, as indicated in Figure \ref{fig:sfr_m}.

\subsection{KCWI Observations}\label{subsec:kcwi}
The galaxies in our sample were observed over the course of eight nights in 2018 -- 2020 using KCWI \citep{Morrissey18} on the Keck II Telescope. The medium slicer and the BL grating with a central wavelength of 4500 \AA{}  were used, resulting in a 16\farcs5 $\times$ 20\farcs0 field of view and a spectral resolution of $R\sim1800$. The typical integration time per pointing was $\sim$ 5 hours and the average seeing was $\sim$ 1\farcs0. The KCWI Data Reduction Pipeline\footnote{\url{https://github.com/Keck-DataReductionPipelines/KcwiDRP}} was used to reduce individual cubes, and the various cubes constructed from exposures at different position angles were combined and drizzled onto a common grid (0\farcs3 $\times$ 0\farcs3) using custom-built Python software as described in \citet{Chen21}.  Briefly, the reduction steps include overscan subtraction, cosmic ray removal, scattered light subtraction, wavelength calibration, flat-fielding, sky-subtraction, cube generation, differential atmospheric refraction correction, and flux calibration. We also used median filtering to remove the low frequency background structures. Finally, multi-band images from 3D-HST survey (described in Section \ref{subsec:3dhst}) were stacked using inverse variance weighting, and this combined image was cross-correlated with the KCWI continuum image to calculate the alignment offset, which was used for the astrometric correction of the KCWI data. We refer readers to \citet{Chen21} for more details.

\begin{figure*}[t]
    \centering
    \includegraphics[width=\textwidth]{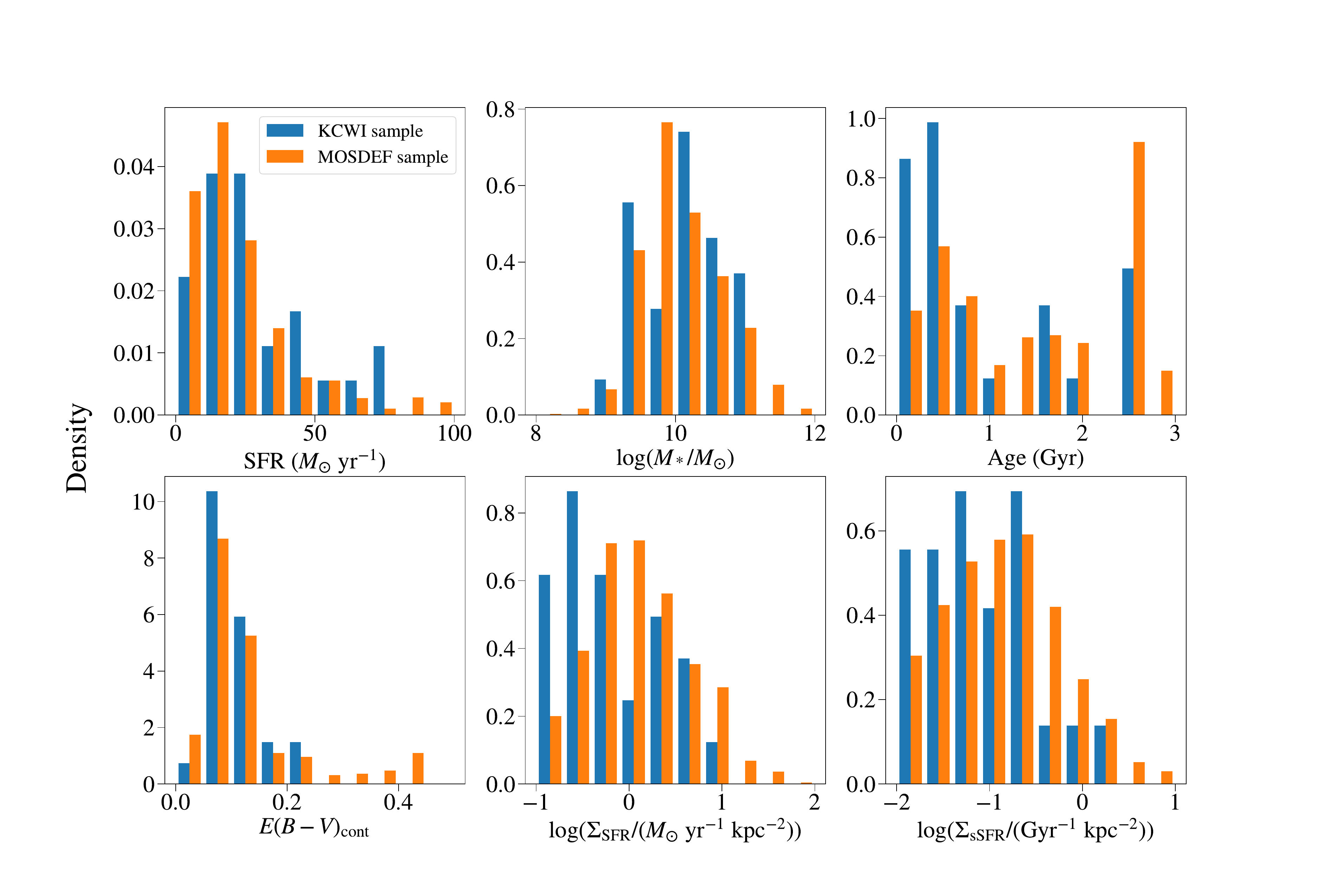}
    \caption{Density histograms of physical quantities of the KCWI sample and the MOSDEF galaxies at the same redshift ($2.0 < z < 3.2$).}
    \label{fig:distribution}
\end{figure*}

\subsection{2D Ly$\alpha$ images} \label{subsec:2dimg}
The 2D \lya{} images of each galaxy were calculated from the 3D data cubes using a method similar to that described in \citet{Erb18}. Briefly:

First, the fluxes in the central 9 $\times$ 9 pixels (2\farcs7 $\times$ 2\farcs7) of the reduced cubes were summed at each wavelength point to produce 1D spectra. For each wavelength point, the flux uncertainty is calculated by summing the uncertainty per pixel in quadrature. The continuum levels on the red and blue sides of the \lya{} line, ${\rm c_{red,spec}}$ and ${\rm c_{blue,spec}}$, were calculated by averaging the spectrum in two windows spanning rest-frame wavelengths of 1269 -- 1279 \AA, and 1160 -- 1180 \AA, respectively. The uncertainties of the red-side and blue-side continuum levels were calculated by summing the uncertainty of the 1D spectrum in quadrature within the two windows. The continuum level at \lya, ${\rm c_{Ly\alpha, spec}}$, was calculated as the average of the blue and red side continuum levels \citep{Kornei10}, while its uncertainty was determined by summing the red-side and blue-side continuum uncertainties in quadrature.

Second, two-dimensional images of \lya{} (${\rm I_{Ly\alpha}}$) and the red side continuum (${\rm I_{cont, red}}$) were extracted from the 3D data cube by collapsing the cube along the dispersion axis and summing over the wavelength regions 1210 -- 1220 \AA{} and 1269 -- 1279 \AA, respectively. The variance cube was summed along the dispersion axis within these two windows to obtain the 2D variance images. The square roots of the variance images were calculated as the uncertainty images.

Lastly, the continuum image underlying \lya{} (${\rm I_{cont, Ly\alpha}}$) and the ``\lya{} only'' image (${\rm I_{Ly\alpha, only}}$) were calculated as follows:
\begin{equation}
    {\rm I_{cont, Ly\alpha} = \frac{c_{Ly\alpha, spec}}{c_{red, spec}}\times I_{cont, red}}
\end{equation}
\begin{equation}
    {\rm I_{Ly\alpha, only} = I_{Ly\alpha} - I_{cont, Ly\alpha}}
\end{equation}
The uncertainties of these two images were calculated following a similar methodology to that described above.

\subsection{3D-HST Images}\label{subsec:3dhst}
We used the publicly-available multi-band (F125W, F140W, F160W, F606W, and F814W) images that were compiled by the 3D-HST grism survey team \citep{Grogin11, Koekemoer11, Skelton14}. The HST images were drizzled to a 0.06 arcsec pixel$^{-1}$ scale and PSF-smoothed to the same 0.18 arcsec spatial resolution as the F160W data. We mosaiced the images from different filters using inverse variance weighting, and this combined image was used for the astrometric correction of the KCWI data and the calculation of the point spread function (PSF) of the KCWI observations.

\subsection{Physical Properties}\label{subsec:sed}
The SED parameters (\mstar, SFR, age, continuum reddening \ebvcont) of the KCWI sample were derived using SED fitting as described in \citet{Reddy15}. We assumed constant star-formation histories, the Bruzual and Charlot \citep[BC03;][]{Bruzual03} stellar population synthesis models at 0.2 $Z_{\sun}$, a Chabrier initial mass function \citep{Chabrier03}, and the SMC attenuation curve \citep{Fitzpatrick90, Gordon03}\footnote{These assumptions are based on previous work \citep{Reddy15,Shivaei15,Reddy18a,Weldon22} that using MOSDEF galaxies.}. The SFR surface density (\sigmasfr) and \sigmasfr{} normalized by stellar mass (\sigmassfr) are calculated as:
\begin{equation}
    {\rm \Sigma_{SFR}} = \frac{\rm SFR}{2\pi r_e^2}
\end{equation}
\begin{equation}
    {\rm \Sigma_{sSFR}} = \frac{\rm SFR}{2\pi r_e^2 M_{*}}
\end{equation}
where SFR and \mstar{} are the star-formation rate and the stellar mass from SED fitting, and $r_e$ is the effective radius from \citet{vanderwel14} which contains half of the total HST/F160W light. In Figure \ref{fig:distribution} we show the histograms of physical quantities of the KCWI sample and the MOSDEF sample. The comparison shows that the KCWI sample is representative of the parent sample from which it was drawn.

\section{Composite Images}\label{sec:composite}
\subsection{\lya{} surface brightness profiles and scale lengths}\label{subsec:lyaprof}

To examine how the halo properties vary with SED parameters, the galaxies were binned according to their SFR, \mstar, ages, \ebvcont, \sigmasfr, and \sigmassfr. For each galaxy, a 9\arcsec{} $\times$ 9\arcsec{} image was extracted from its ``\lya{} only" image. The unweighted average of these images was then used to create a composite image of each subsample. The uncertainty images were summed in quadrature to calculate the uncertainty of the composite image. The surface brightness profiles were calculated based on the stacked images in annuli with radii r = 0 to 4\farcs5 and a width of 0\farcs15. We used the Python package \texttt{photutils}\footnote{\url{https://photutils.readthedocs.io/en/stable/}} to calculate the surface brightness flux density and its uncertainty.
\begin{deluxetable}{cccc}
\tablecolumns{3}
\tablecaption{Scale lengths $r_n$ in pkpc.\label{tab:scalelength}}
\tablehead{
\colhead{Parameters} & \colhead{Threshold} & \colhead{Low bin} & \colhead{High bin}}
\startdata
SFR & $22\ M_{\odot}\ {\rm yr^{-1}}$  & 15.5$\pm$3.7 & 10.9$\pm$0.4 \\
\mstar & $10^{10.1}\ M_{\odot}$ & 10.6$\pm$0.4 & 12.6$\pm$0.7 \\
Age & $\rm 0.5\ Gyr$ &8.8$\pm$0.3 & 15.5$\pm$0.9 \\
\ebvcont & $0.09$ & 9.2$\pm$0.4 & 21.6$\pm$1.5 \\
\sigmasfr & $0.78\ M_{\odot}\ {\rm yr^{-1}\ kpc^{-2}}$ & 21.9$\pm$3.2 & 8.5$\pm$0.5 \\
\sigmassfr & $0.035\ {\rm Gyr^{-1}\ kpc^{-2}}$& 17.4$\pm$1.1 & 8.7$\pm$0.4 \\
\enddata
\end{deluxetable}

The surface brightness profile of the \lya{} halo is usually described by a decreasing exponential model \citep{Steidel11}:

\begin{equation}
    S(r)=C_ne^{-r/r_n}
	\label{eq:scalelength}
\end{equation}
where $C_n$ is a constant and $r_n$ is the scale length. However, since the profiles are not monotonically decreasing, we fit the surface brightness profiles with equation (\ref{eq:scalelength}) convolved with the seeing PSF at radii spanning from the radius at which \lya{} peaks, out to 4\farcs5. The seeing PSF for each subsample was calculated using the full width at half-maximum (FWHM) of gaussian fits to the stacked HST and KCWI continuum images, i.e., $\rm FWHM_{PSF} = \sqrt{FWHM_{KCWI}^2 - FWHM_{HST}^2}$. The average redshift of each subsample was then used to convert the scale length in angular size to physical distance. We note that the effect of the PSF is marginal since the \lya{} emission is extended and the fitting range is large compared to the FWHM of the PSF. The scale lengths $r_n$ in proper kpc (pkpc) for different subsamples are listed in Table \ref{tab:scalelength}.

\begin{figure*}
\includegraphics[width=\textwidth]{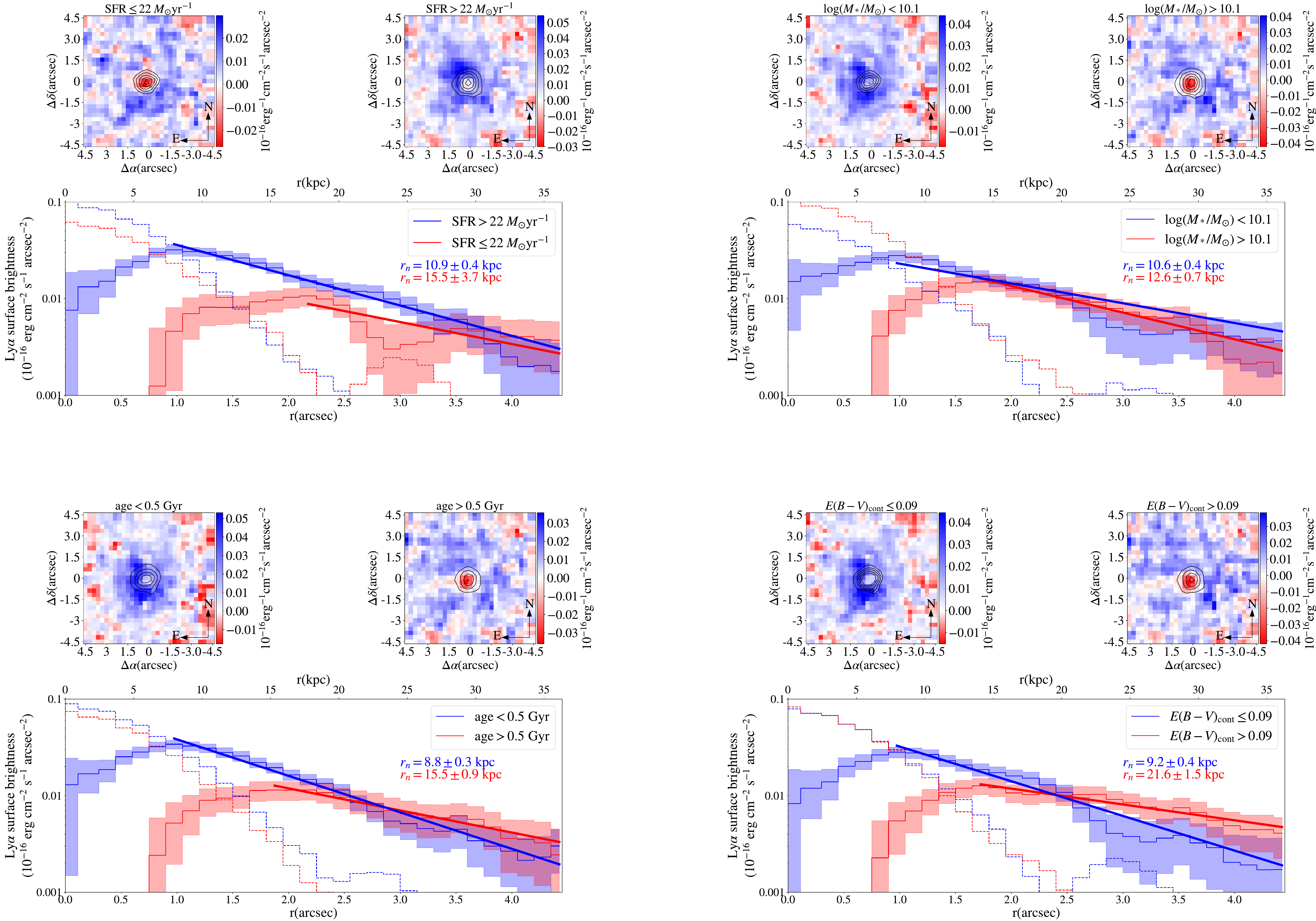}
\caption{Top in each panel: Composite \lya{} images of SFR, \mstar, age, and \ebvcont{} subsamples. There are an even number of galaxies in each subsample. The black contours indicate the surface brightness of the composite continuum images, with the lowest level at ${\rm 2.4\times10^{-18}\,erg\,s^{-1}\,cm^{-2}\,arcsec^{-2}}$. Bottom in each panel: Surface brightness profiles of composite \lya{} images. The shaded regions indicate the 1$\sigma$ error of the mean. The dashed lines indicate the surface brightness profiles of the continuum images. The straight lines indicate the best fitting lines. The physical radius is calculated based on the average redshift of the sample ($z$ = 2.41). $r_n$ indicates the scale length of different subsample.\label{fig:profile1}}
\end{figure*}

\begin{figure*}
\includegraphics[width=\textwidth]{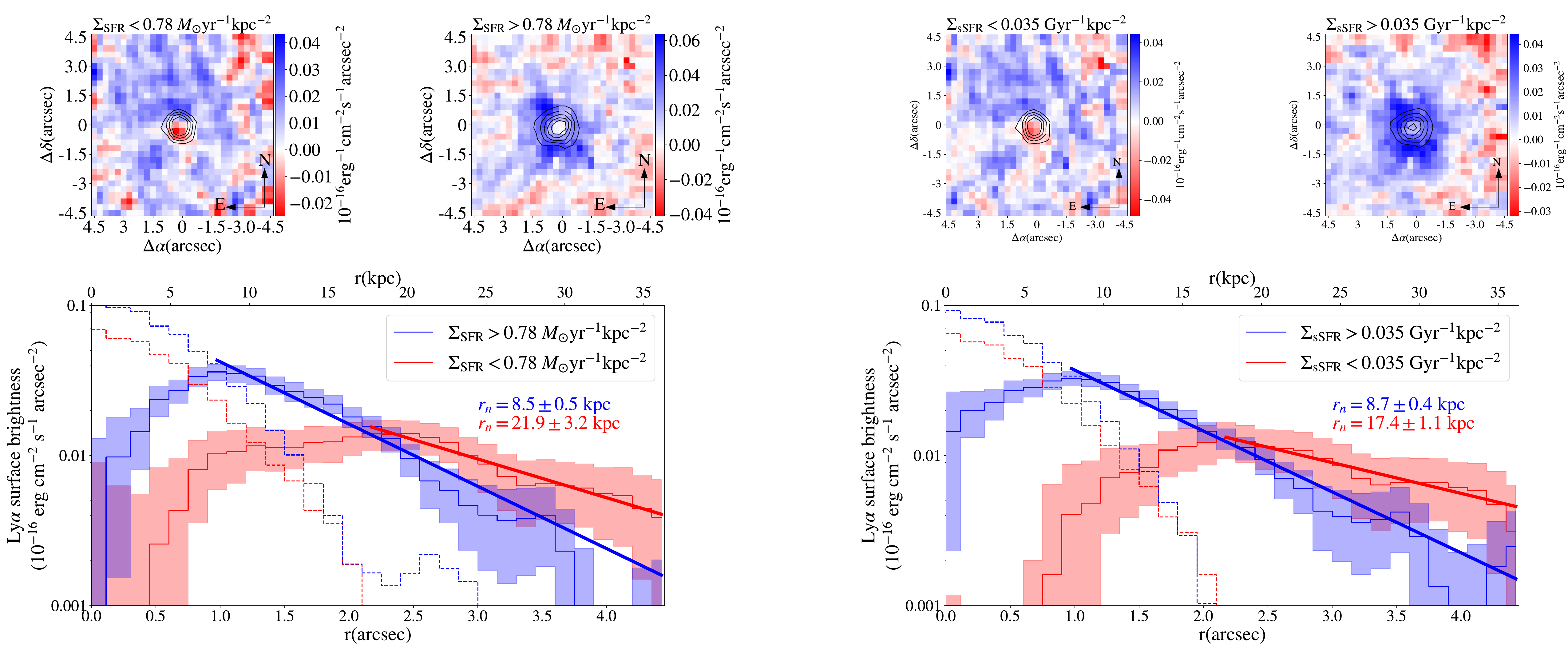}
\caption{Same as Figure \ref{fig:profile1} but for \sigmasfr{} and \sigmassfr{} subsamples.\label{fig:profile2}}
\end{figure*}

Figure \ref{fig:profile1} and \ref{fig:profile2} show the composite profiles in two bins of SFR, stellar mass, age, \ebvcont, \sigmasfr{} and \sigmassfr, respectively. The profiles have a non-monotonic shape with a peak at $r \sim 1$ arcsec for high SFR, low \mstar, young ages, low reddening, and high \sigmassfr{} galaxies. On the other hand, the profiles for low SFR, high \mstar, high reddening, old ages, and low \sigmassfr{} exhibit a deficiency of \lya{} within a $\sim$ 0.7 arcsec radius. The former subsamples also have smaller scale lengths. For the SFR and stellar mass subsamples, the differences in the scale lengths between the low and high bins are $\sim$ 1.2 $\sigma$ and $\sim$ 2.5 $\sigma$, respectively, the latter being marginally significant. The differences in the scale lengths of age, \ebvcont{} and \sigmassfr{} subsamples are significant ($> 7 \sigma$). Meanwhile, deficits of \lya{} emission are indicated in both \sigmasfr{} subsamples, while the scale lengths of the halos are significantly different at the $> 4 \sigma$ level. The implications of these results are discussed in Section \ref{subsec:scalelength}.

\subsection{Down-the-barrel \lya{} fractions}\label{subsec:dtb}
In Table \ref{tab:dtb}, we present the down-the-barrel fractions (or upper limits) of \lya{} emission. This quantity tells us the fraction of \lya{} escaping along the same lines of sight that intersect with the non-resonant continuum emission. The down-the-barrel fraction \fdtb{} is calculated as
\begin{equation}
    f_{\rm dtb} = L_{\rm Ly\alpha,c}/L_{\rm Ly\alpha,total}
\end{equation}
where $L_{\rm Ly\alpha,c}$ is the \lya{} emission within an aperture of diameter equal to the FWHM of the continuum emission, and $L_{\rm Ly\alpha,total}$ is the \lya{} emission within an aperture that yields the highest S/N measurement of \lya. All the fractions are less than 10\%, indicating that the vast majority of \lya{} emission is resonantly scattered far from the continuum emission region. For all SED properties examined here, the subsamples with larger scale lengths show absorption in their central regions and the differences in the down-the-barrel \lya{} fraction between subsamples is significant ($> 3 \sigma$). These results are discussed in Section \ref{subsec:deficit_of_lya}.

\begin{deluxetable}{cccc}
\tablecaption{Down-the-barrel \lya{} emission fraction\label{tab:dtb}}
\tablehead{\colhead{Parameters} & \colhead{Threshold} & \colhead{Low bin} & \colhead{High bin}}
\startdata
SFR & $22\ M_{\odot}\ {\rm yr^{-1}}$ & $<0.05$ & 0.06$\pm$0.01 \\
\mstar & $10^{10.1}\ M_{\odot}$ & 0.09$\pm$0.01 & $<0.03$ \\
Age & $\rm 0.5\ Gyr$ & 0.09$\pm$0.01 & $<0.03$ \\
\ebvcont & $0.09$ & 0.08$\pm$0.01 & $<0.03$ \\
\sigmasfr & $0.78\ M_{\odot}\ {\rm yr^{-1}\ kpc^{-2}}$ & 0.01$\pm$0.01 & 0.10$\pm$0.01 \\
\sigmassfr & $0.035\ {\rm Gyr^{-1}\ kpc^{-2}}$ & $<0.024$ & 0.10$\pm$0.01 \\
\enddata
\end{deluxetable}

\section{Dust Attenuation of \lya}\label{sec:individual}
\subsection{\lya{} escape fraction and equivalent width}

A number of studies suggest that Lyman continuum (LyC) photons generally escape through the same low-column-density channels in the ISM as \lya{} photons \citep{Gnedin08, Zackrisson13, Trainor15, Dijkstra16, Ma16, Reddy16, Steidel18, Kimm19, Ma20, Kakiichi21}. Thus studying the \lya{} escape fraction is useful for understanding the escape of LyC photons, an important factor in cosmic reionization \citep{Miralda00}. Since \lya{} photons can also resonantly scatter through the CGM and suffer preferential dust attenuation, the \lya{} escape fraction can reveal information on dust in the CGM.

Here, we computed the effective attenuation of the \lya{} line using the following procedure. \citet{reddy20} showed that the nebular dust attenuation curve derived for MOSDEF galaxies at $z\sim2$ is similar to the Galactic extinction curve \citep{cardelli89}. Using H$\alpha$ and H$\beta$ emission lines, corrected for Balmer absorption, we computed $E(B-V)_{\rm neb}$ and the intrinsic (dust-corrected, assuming the Galactic extinction curve) H$\alpha$ luminosities. The \lya{} (continuum) luminosities were calculated by finding the aperture that yields the highest S/N measurement of \lya{} (continuum). A commonly used intrinsic flux ratio is $F_{\rm Ly\alpha}/F_{\rm H\alpha} = 8.7$ under the assumption of case B recombination and $T_e=10^4\,$K \citep{Brocklehurst71}. However, as shown by \citet{Reddy22}, an intrinsic flux ratio of 9.3 is more appropriate for galaxies in the MOSDEF survey, thus we adopt this value. The \lya{} escape fraction $f_{\rm esc}$ is defined as
\begin{equation}
    f_{\rm esc}=\frac{L_{\rm Ly\alpha, obs}}{L_{\rm Ly\alpha, int}}=\frac{L_{\rm Ly\alpha, obs}}{9.3L_{\rm H\alpha, int}}
\end{equation}
where $L_{\rm Ly\alpha, obs}$ and $L_{\rm Ly\alpha, int}$ are the total observed and intrinsic \lya{} luminosities, and $L_{\rm H\alpha, int}$ is the intrinsic H$\alpha$ luminosity.

\begin{figure}
\includegraphics[width=\columnwidth]{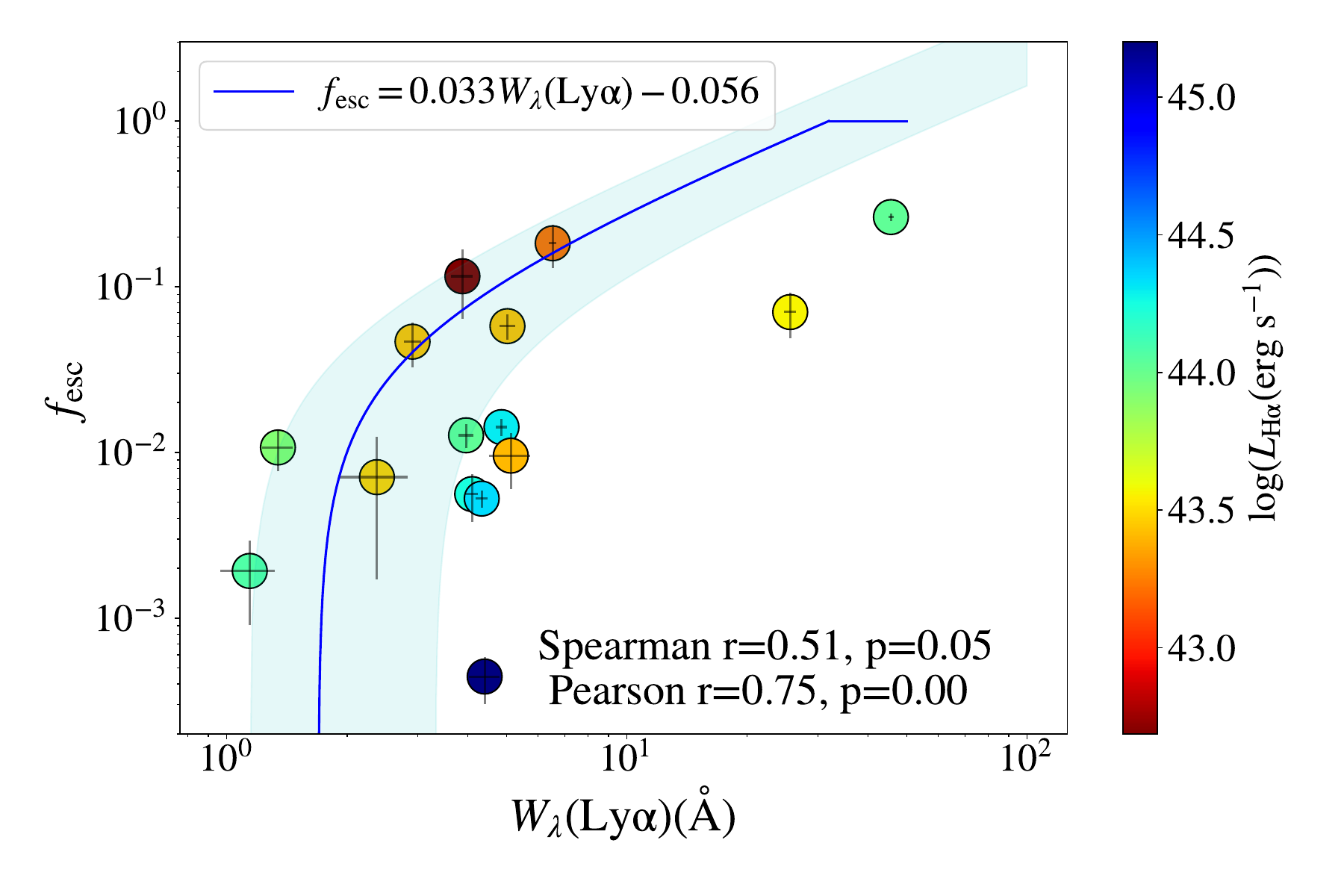}
\caption{\lya{} escape fraction versus \lya{} equivalent width in rest-frame. The shaded region indicates the 1$\sigma$ confidence interval. The data points are colored by their $\rm H\alpha$ luminosities. \label{fig:ew_fesc}}
\end{figure}

The \lya{} escape fraction $f_{\rm esc}$ as a function of rest-frame equivalent width $W_{\rm \lambda}{\rm (Ly\alpha)}$ (integrated over the entire \lya{} halo) is presented in Figure \ref{fig:ew_fesc}. The rest-frame \lya{} equivalent width was calculated using the observed \lya{} and continuum luminosities. Note that for the following discussion, 11 galaxies that are close to the edge of field of view or their neighbors are excluded since their \lya{} luminosities could be underestimated due to the small field of view, or overestimated due to the contamination of their neighbors. A linear correlation is found between $f_{\rm esc}$ and $W_{\rm \lambda}{\rm (Ly\alpha)}$ with a Pearson correlation coefficient $r = 0.57 $ and a probability of null correlation of $p = 0.03$. This is consistent with the result of \citet{Trainor15} and \citet{Reddy22} who found that $W_{\rm \lambda}{\rm (Ly\alpha)}$ correlates with $f_{\rm esc}$. Since the intrinsic \lya{} luminosity $L{\rm (Ly\alpha)_{int}}$ scales with the intrinsic $\rm H\alpha$ luminosity $L{\rm (H\alpha)_{int}}$, and $W_{\rm \lambda}{\rm (Ly\alpha)}$ is proportional to the ratio of the observed \lya{} luminosity to observed UV luminosity $L{\rm (UV)_{obs}}$,
\begin{equation}\label{eq:fesctoew}
    \frac{f_{\rm esc}}{W_{\lambda}{\rm (Ly\alpha)}} \propto \frac{L{\rm (UV)_{obs}}}{L{\rm (H\alpha)_{int}}} = \frac{L{\rm (UV)_{int}}}{L{\rm (H\alpha)_{int}}}\times10^{-0.4k({\rm UV})E(B-V)}
\end{equation}
The scatter in the linear relation between $f_{\rm esc}$ and $W{\rm _{\lambda}}{\rm (Ly\alpha)}$ stems from differences in \ebv{} and the variation in H$\alpha$-to-UV ratio. For galaxies with higher attenuation, $W_{\rm \lambda}{\rm (Ly\alpha)}$ would be larger relative to $f_{\rm esc}$. On the other hand, \citet{Fetherolf21} and \citet{Rezaee22} found that galaxies with higher H$\alpha$ luminosities (higher instantaneous SFR) have higher $\rm H\alpha$-to-UV ratios. In Figure \ref{fig:ew_fesc}, galaxies with relatively lower $f_{\rm esc}$ have larger H$\alpha$ luminosities, and hence have lower $f_{\rm esc}/W_{\lambda}{\rm (Ly\alpha)}$.

\subsection{Enhanced attenuation of \lya{} photons}

\begin{figure}
\includegraphics[width=9.5cm]{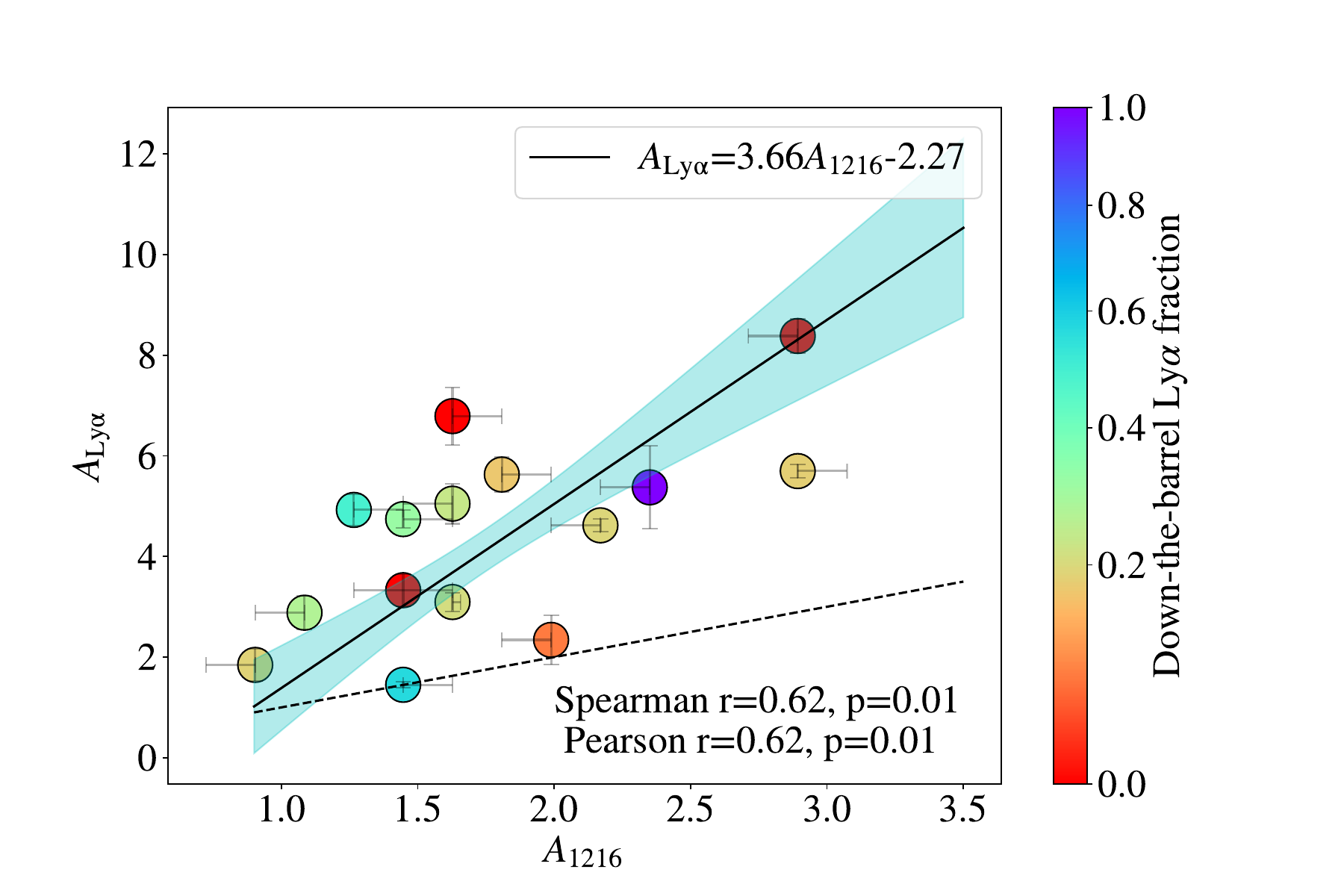}
\caption{$A_{\rm Ly\alpha}$ as a function of $A_{\rm 1216}$. The solid line indicates the best-fitting line while the dashed line indicates the identity line. The shaded region indicates the 1$\sigma$ confidence interval. The data points are colored by their down-the-barrel \lya{} fraction.\label{fig:AlyavsAcont}}
\end{figure}

Due to their resonant scattering, \lya{} photons are more likely to be destroyed by dust grains in the CGM relative to the non-resonant UV continuum photons. We can test for this effect using our data. Figure \ref{fig:AlyavsAcont} shows the attenuation of \lya{} ($A_{\rm Ly\alpha}$) versus the attenuation of continuum flux at 1216 \AA{} ($A_{\rm 1216}$). The attenuation of \lya{} photons can be derived from the \lya{} escape fraction, i.e., ${A_{\rm Ly\alpha}} = -2.5{\rm log}(f_{{\rm esc}})$. The attenuation of continuum flux at 1216 \AA{} was calculated using the SMC attenuation curve and \ebvcont. The data points are color coded by the down-the-barrel \lya{} fraction (the ratio of \lya{} flux spatially coincident with the continuum to the total \lya{} flux). $A_{\rm Ly\alpha}$ is found to be positively correlated with $A_{\rm 1216 }$ and $A_{\rm Ly\alpha}$ is greater than $A_{\rm 1216}$ for dustier galaxies. This result shows that the resonant scattering of \lya{} results in a higher effective attenuation of \lya{} photons relative to the non-resonant UV continuum photons.

\begin{figure*}
\includegraphics[width=\textwidth]{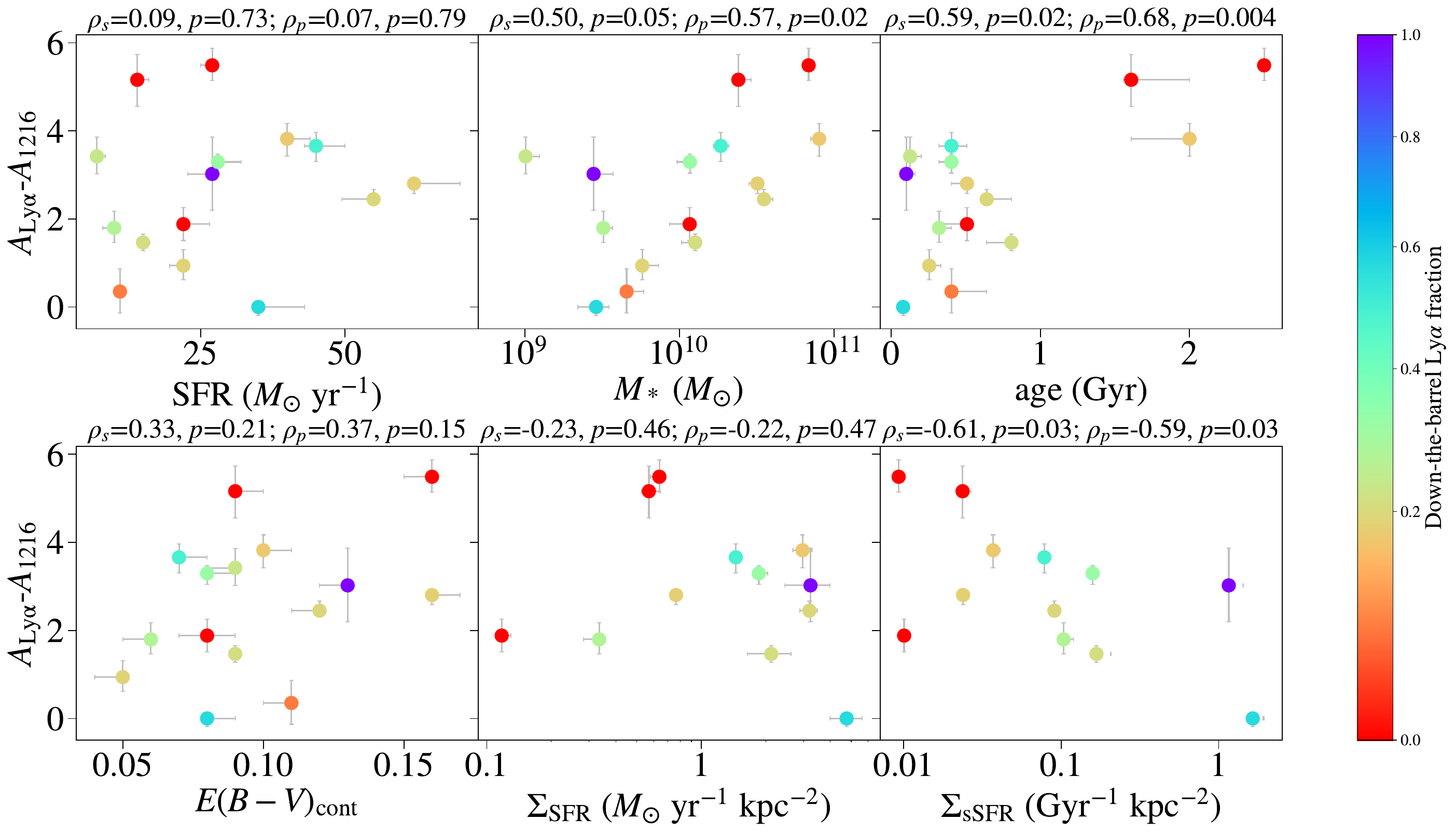}
\caption{$A_{\rm Ly\alpha}$ - $A_{\rm 1216}$ as a function of different physical properties. Spearman and Pearson correlation coefficients and their p-values are shown on the top of each grid. The data points are colored by their down-the-barrel \lya{} fraction. The only significant correlations are with \mstar, age, and \sigmassfr.\label{fig:enhancedA}}
\end{figure*}

The enhancement of attenuation for \lya, $\Delta A = {A_{\rm Ly\alpha}} - {A_{\rm 1216}}$, as a function of different physical properties is shown in Figure \ref{fig:enhancedA}. The difference in attenuation is correlated with stellar mass and age, while anticorrelated with \sigmassfr. A weak correlation is found for \ebvcont. No correlation is found for SFR and \sigmasfr. The implication of these results are discussed in Section \ref{subsec:dust}.

\section{Discussion}\label{sec:discussion}
In this section, we discuss how the down-the-barrel \lya{} fraction (Section \ref{subsec:deficit_of_lya}), scale lengths of \lya{} halos (Section \ref{subsec:scalelength}), and dust in the CGM depend on physical properties of their host galaxies (Section \ref{subsec:dust}). These relations are important for understanding the escape of \lya{} photons and the properties of the CGM.

\subsection{Down-the-barrel \lya{} fraction}\label{subsec:deficit_of_lya} 
There are two ways for \lya{} photons to escape from the CGM of a galaxy: one is to escape after multiple resonant scatterings \citep{Laursen07}; the other is to escape through the low-column-density channels that are created by strong stellar winds from massive stars and/or supernovae \citep{Gnedin08, Zackrisson13, Ma16, Sharma17}. The latter is also thought to be a mechanism through which LyC photons escape \citep{Gnedin08, Ma16, Kimm19, Ma20, Kakiichi21}. Thus the dependence of down-the-barrel \lya{} fractions (\fdtb) on physical properties can reveal how these low-column-density channels are regulated.

In Table \ref{tab:dtb}, we list the down-the-barrel \lya{} fractions of different subsamples. Upper limits of \fdtb{} are obtained for galaxies with lower SFR and \sigmassfr, larger stellar mass and \ebvcont, and older ages. A deficit of \lya{} is also observed in the center of stack images of low \sigmasfr{} galaxies (Figure \ref{fig:profile1} and \ref{fig:profile2}). A deficit of \lya{} is also observed for the ``\lya{} Abs" sample (i.e., those galaxies with the down-the-barrel \lya{} in net absorption) of \citet{Steidel11} and in the non-LAE realizations of \citet{Momose14, Momose16}. On the other hand, the \lya{} surface brightness profiles of the remaining subsamples are suppressed in the central regions. This effect has also been reproduced by the simulation of \citet{Laursen09}, in which they found that this suppression is caused by the inclusion of dust. Equation (2) of \citet{Steidel11} provided a model that explains the deficit and suppression of \lya: \lya{} photons created in the central region are destroyed or scattered to outer regions because of the high covering fraction of neutral gas at small impact parameters and near the systemic redshift of the galaxy.

We find that galaxies with higher SFRs also have higher \fdtb. This is consistent with a scenario in which low-column-density channels are caused by stellar feedback that is prevalent in galaxies with higher SFRs. However, the difference between \fdtb{} of the high SFR subsample and the upper limit of \fdtb{} of low SFR subsample is $\sim 1 \sigma$. This insignificant difference may be due to the fact that galaxies with higher SFRs are also more gas- (and dust-) rich \citep{Reddy10, Reddy15, Dominguez13, Sanders22}, such that it may be more difficult for stellar feedback to puncture channels through the ISM/CGM of these galaxies. Furthermore, high SFR galaxies are generally more massive (Fig. \ref{fig:sfr_m}), and in the following analysis, we will show that the high gravitational potential associated with high-stellar-mass galaxies can impede the creation of low-column-density channels. Thus SFR alone may not be a good indicator of \fdtb{} and hence LyC escape.

A commonly used indicator of LyC escape is \sigmasfr{} \citep{Alexandroff15, Sharma16, Naidu20, Flury22}, as it is a proxy of star formation feedback and potentially the creation of low-column-density channels \citep{Heckman01}. A significant difference in \fdtb{} is found between high and low \sigmasfr{} galaxies, which is expected since high \sigmasfr{} galaxies are believed to be quite efficient at creating low-column-density channels in the ISM \citep{Sharma16, Sharma17, Verhamme17, Cen2020, Naidu20}. Yet there may be other factors that influence the down-the-barrel fraction of \lya. In particular, \citet{Reddy22} highlight the potential importance of gravitational potential in influencing the porosity of the ISM and the leakage of \lya{} and LyC photons. Stellar mass is a rough proxy for dynamical mass or gravitational potential \citep{Price20} and we do find that more massive galaxies exhibit lower \fdtb, suggesting that the gravitational potential may play a role in regulating the escape of \lya. Therefore we examined ${\rm \Sigma_{SFR}}/M_*$ to ascertain whether galaxies at a fixed \sigmasfr{} but lower $M_*$ have a larger down-the-barrel \lya{} escape fraction. We find that the high \sigmassfr{} bin has a larger down-the-barrel fraction than the low \sigmassfr{} bin. When binned solely by \sigmasfr, the difference in \fdtb{} between subsamples is smaller than when binned by \sigmassfr. Therefore, galaxies with high and low \fdtb{} can be more effectively separated by \sigmassfr{} (Figure \ref{fig:enhancedA}). This suggests that gravitational potential may be an important factor in the porosity of the ISM and the leakage of \lya{} photons, consistent with the analysis of \citet{Reddy22}.

We also examine the dependence of \fdtb{} on reddening, finding that galaxies with lower \ebvcont{} show higher \fdtb. This result is consistent with \citet{Reddy16}, who found that dustier galaxies have larger neutral gas covering fractions. We also find that \fdtb{} anti-correlates with the age of galaxies, which is expected given that young galaxies are less massive and less dusty.

\subsection{Scale lengths of \lya{} halos}\label{subsec:scalelength}
The sizes of \lya{} halos are typically parameterized by their exponential scale lengths. The relations between the scale lengths and other physical quantities are of interest since they reveal the amount and distribution of gas and dust in the CGM. In Table \ref{tab:scalelength}, we report the scale lengths of the profiles of the various subsamples, which are found to vary from $\sim$ 8 to 25 pkpc. These results are consistent with both \citet{Steidel11} (UV continuum selected galaxies; $\sim$ 25 pkpc) and \citet{Momose16} (\lya{} emitters; $\sim$ 10 pkpc), and are slightly larger than the scale lengths found by \citet{Xue17} (UV-continuum faint \lya{} emitters; $\sim$ 6 pkpc). The sample used in \citet{Xue17} contains galaxies with $z \sim 3.78$, whose UV continuum sizes are smaller than those in our sample. Given the relation between \lya{} halo scale length and UV continuum scale length \citep{Wisptzki16, Leclercq17}, it is reasonable to expect that our KCWI sample has longer \lya{} scale length than the sample used in \citet{Xue17}.

Comparing Table \ref{tab:dtb} and Table \ref{tab:scalelength}, we find that subsamples with higher \fdtb{} have shorter scale lengths. This relation indicates that the number of low-column-density channels might regulate the scale length of \lya{} halos. A natural explanation is that with more low-column-density channels, the fraction of \lya{} photons that are scattered into the CGM is lower. This effect would enhance the \lya{} surface brightness profile in the inner region and reduce it in the outer region, thus resulting in a shorter scale length for the extended \lya{} emission.

For our subsamples, scale lengths of high SFR galaxies are found to be slightly smaller ($\sim$ 1 $\sigma$) than those of low SFR galaxies. \citet{Wisptzki16} and \citet{Leclercq17} also found that for LAEs, $M_{\rm UV}$, which is an indicator of SFR, is correlated with the scale length. However, in other studies of LAEs, \citet{Matsuda12} found no correlation between scale length and central (within 1 arcsec) UV luminosity. Moreover, \citet{Feldmeier13} showed that UV bright galaxies have slightly larger \lya{} halos. These mixed results could arise from the fact that star-forming galaxies are also gas- and dust-rich, so there may not be a direct correlation between SFR and the scale length in the same way that we found no correlation between SFR and \fdtb{} (Section \ref{subsec:dtb}).

Table \ref{tab:scalelength} indicates that scale lengths increase with reddening (or dustiness). This result would be expected if galaxies with more dust contain more gas and therefore have higher gas covering fractions \citep{Reddy16}. In this case, a larger fraction of \lya{} photons scatter away from the continuum regions and either escape or are destroyed by dust at larger radii. The resonant scattering would flatten the \lya{} surface brightness profile, resulting in a longer scale length.

High mass galaxies in the KCWI sample are found to have slightly larger \lya{} halos likely because they have both higher SFRs and are dustier. In Section \ref{subsec:deficit_of_lya} we also show that massive galaxies have stronger gravitational potential, which impedes the creation of low-column-density channels. \citet{Reddy22} also found that \lya{} equivalent width anti-correlates with stellar mass, suggesting lower gas covering fractions in less massive galaxies. Combining the effect of stellar mass and dustiness, it is not surprising that younger galaxies have shorter scale lengths given that they are less dusty and less massive. We also find that the younger galaxies have higher \sigmassfr{} (Pearson $r=-0.772$, $p$-value $<0.001$), and hence have a larger fraction of \lya{} emission coming from the continuum region, and a smaller fraction that is resonantly scattered to large radii. Thus, these younger galaxies have shorter halo scale lengths.

Unlike SFR, the impact of SFR surface density on the scale length is significant. The high \sigmasfr{} subsample exhibits the smallest scale length ($<$ 10 pkpc) of any other subsamples while the scale length of the low \sigmasfr{} subsample is the largest among all the subsamples. Galaxies with higher \sigmasfr{} are more compact, thus their gas distribution may also be more compact. Their \lya{} surface brightness profiles are then expected to be decreasing steeply. In Section \ref{subsec:deficit_of_lya} we discussed that high \sigmassfr{} plays an important role in creating low-column-density channels in the ISM. Those channels would ease the escape of \lya{} photons at smaller radius and therefore result in a smaller scale length.

\subsection{Dust in the CGM}\label{subsec:dust}

Due to their resonant nature, \lya{} photons may suffer varying degrees of attenuation relative to the stellar continuum.  In particular, a larger number of resonant scatterings in a dusty medium results in a higher probability for \lya{} photons to be absorbed by dust, resulting in an increase in the attenuation of \lya. On the other hand, \lya{} photons exiting down the barrel of the galaxy are likely undergoing fewer resonant scatterings and escaping the ISM through channels of low gas and dust column densities.  In this case, these \lya{} photons may not suffer much attenuation \citep[e.g.,][]{Trainor15, Reddy22}. \citet{Scarlata09} showed that observed high \lya/H$\alpha$ and H$\alpha$/H$\beta$ line ratios can be reproduced by a clumpy dust distribution, implying the existence of low-column-density sightlines. To study the dust content of the CGM, we focus on the attenuation of \lya{} photons that escape from the halo. When assuming that all the down-the-barrel emission comes from the low-column-density sightlines and is not significantly attenuated, the escape fraction of the halo \lya{} photons can be defined as
\begin{equation}
f_{\rm halo} = (L_{\rm obs}-L_{\rm dtb})/(L_{\rm int}-L_{\rm dtb})
\end{equation}
where $L{\rm _{obs}}$, $L{\rm _{dtb}}$ and $L{\rm _{int}}$ represent the total observed, down-the-barrel, and intrinsic \lya{} luminosity, respectively. Because most of the \lya{} emission comes from the halo ($>90\%$), $L{\rm _{dtb}}$ is typically much smaller than $L{\rm _{obs}}$ and $L{\rm _{int}}$. Therefore, we can make the approximation that:
\begin{equation}
    f_{\rm halo} \approx L_{\rm obs}/L_{\rm int} = f_{\rm esc}.
\end{equation}
Since $f_{\rm esc}=10^{-0.4A_{\rm Ly\alpha}}$, $A_{\rm Ly\alpha}$ can be used to describe the attenuation of halo \lya{} photons.

\citet{Reddy16} found that galaxies with higher \ebv{} have a larger covering fraction of optically-thick HI gas and therefore a larger effective attenuation for \lya. It would also not be unreasonable to expect that more reddened galaxies also have more dust in their CGM. This could explain why $\Delta A$ correlates with stellar mass and age, since massive and older galaxies have higher \ebv. However, we only find a marginal correlation between \ebvcont{} and $\Delta A$. The marginal correlation may arise from the fact that stellar feedback would expel dust into the CGM and therefore reduce \ebvcont. As a result, galaxies with low \ebvcont{} may also have substantial dust in their CGM.

In our sample, we find no relation between \ebvcont{} and \sigmassfr, indicating that higher \sigmassfr galaxies may not have more dust in their CGM. Given that no relation is found between \sigmasfr{} and $\Delta A$, we conclude that the anticorrelation between \sigmasfr{} and $\Delta A$ is due to the relation between stellar mass and $\Delta A$.

\section{Summary}\label{sec:conclusion}
In this paper, we utilize KCWI to observe the \lya{} halos of 27 galaxies with spectroscopic redshifts. We study the relations between halo properties and physical quantities. The dust content in the CGM is also discussed. The major results of our paper are summarized as follows:

1. We find extended \lya{} halos in the stacks of all subsamples (Figure \ref{fig:profile1} and \ref{fig:profile2}). A deficit of \lya{} is detected in the center of galaxies with lower SFR and \sigmassfr, higher stellar masses, older ages and higher \ebvcont. Both \sigmasfr{} subsamples show a deficit of \lya{} in their center. For the rest of subsamples, the \lya{} surface brightness densities are suppressed in the central region, which has not been seen in past studies of LAEs.

2. We investigate the down-the-barrel \lya{} fraction and the scale length of \lya{} halos as a function of various physical quantities (Table \ref{tab:scalelength} and \ref{tab:dtb}). Down-the-barrel \lya{} fraction correlates with SFR, \sigmasfr, and \sigmassfr, and anti-correlates with stellar mass, age, and \ebvcont. On the other hand, scale length correlates with stellar mass, age, and \ebvcont, and anti-correlates with SFR, \sigmasfr, and \sigmassfr.

3. The effective attenuation of \lya{} is higher than the attenuation of UV continuum photons at the same wavelength (Figure \ref{fig:AlyavsAcont}). The enhancement of attenuation correlates with stellar mass, age, and \ebvcont{} and anti-correlates with \sigmassfr. No correlation is found for SFR and \sigmasfr{} (Figure \ref{fig:enhancedA}).

In this paper, we show that \lya{} halo properties are regulated by the neutral gas covering fraction (which is indicated by the \lya{} down-the-barrel fraction). This covering fraction is affected by factors such as the amount of gas and dust, and stellar feedback. We also investigate the role of gravitational potential in affecting the intensity of stellar feedback and regulating the \lya{} halo properties. In addition, by examining the enhancement of \lya{} attenuation, we find that more reddened galaxies have more dust in their CGM.

In this paper, we present an analysis of the variation in \lya{} halo profiles with commonly-determined physical properties of galaxies, including stellar mass, SFR, reddening, and SFR surface density. There are other properties of galaxies that are believed to correlate with \lya{} (and LyC) escape, including the [O\,{\footnotesize{III}}]/[O\,{\footnotesize{II}}] ratio and interstellar absorption line equivalent widths. A future study will focus on these additional properties to shed further light on the mechanisms for \lya{} and LyC escape.

%% Also note that the akcnowlodgment environment does not support long amounts of text. If you have a lot of people and institutions to acknowledge, do not use this command. Instead, create a new \section{Acknowledgments}.
\begin{acknowledgments}
We acknowledge support from NSF AAG grants AST1312780, 1312547, 1312764, and 1313171, grant AR13907 from the Space Telescope Science Institute, and grant NNX16AF54G from the NASA ADAP program. We thank the 3D-HST Collaboration, which provided the spectroscopic and photometric catalogs used to select the MOSDEF targets and derive stellar population parameters. This research made use of Astropy\footnote{\url{https://www.astropy.org/}}, a community-developed core Python package for Astronomy \citep{astropy:2013, astropy:2018, astropy:2022}. The authors wish to recognize and acknowledge the very significant cultural role and reverence that the summit of Maunakea has always had within the indigenous Hawaiian community. We are most fortunate to have the opportunity to conduct observations from this mountain.
\end{acknowledgments}

\bibliography{main}{}
\bibliographystyle{main}



\end{document}